%% file: main.tex
\documentclass[preprint]{aastex63}
\usepackage[utf8]{inputenc}
\usepackage{graphicx} 
\usepackage{amsmath}
\usepackage{appendix}

\begin{document}

\newcommand{\two}{$N\hspace{-0.2em}\ge\hspace{-0.2em}2$ }
\newcommand{\three}{$N\hspace{-0.2em}\ge\hspace{-0.2em}3$ }
\newcommand{\four}{$N\hspace{-0.2em}\ge\hspace{-0.2em}4$ }

\title{Architectures of Planetary Systems III: Excitation of Eccentricities and Inclinations}

\author[0000-0002-4884-7150]{Alex R. Howe}
\affiliation{The Catholic University of America, 620 Michigan Ave., N.E. Washington, DC 20064}
\affiliation{NASA Goddard Space Flight Center, 8800 Greenbelt Rd, Greenbelt, MD 20771, USA}
\affiliation{Center for Research and Exploration in Space Science and Technology, NASA/GSFC, Greenbelt, MD 20771}

\author[0000-0002-7733-4522]{Juliette C. Becker}
\affiliation{Department of Astronomy, University of Wisconsin-Madison, 475 N. Charter Street, Madison, WI 53706, US}


\author{Fred C. Adams}
\affiliation{Department of Physics, University of Michigan, 450 Church St, Ann Arbor, MI 48109}
\affiliation{Department of Astronomy, University of Michigan, Ann Arbor, MI 48109}

\begin{abstract}

The current census of planetary systems displays a wide range of architectures. Extending earlier work, this paper investigates the correlation between our classification framework for these architectures and the distribution of eccentricities and inclinations of planetary orbits using both dispersion and normalized angular momentum deficit (NAMD) metrics. Orbital inclinations prove to be too strongly affected by observational biases to yield meaningful results, but the patterns in eccentricities reveal significant correlations. We find that systems with large gaps between planets are more dynamically excited than closely-spaced systems, based on dispersion in eccentricity and an eccentricity-specific NAMD metric. Systems with detected outer planets also appear dynamically excited, to a degree where our own solar system appears to be dynamically colder than expected. We also predict the existence of a population of highly-inclined long-period planets that is likely to be observed by upcoming astrometric surveys.

\end{abstract}

\section{Introduction}
\label{sec:intro}

The current population of confirmed extrasolar planets stands at over 6000, including over 1000 multiplanet systems. This census has grown large enough that a population-level analyses of planetary system architectures can be informative. We explored this idea in our previous two papers. In \cite{PaperI}, henceforth Paper I, we proposed a classification framework for system architectures and classified all planetary systems that were well enough characterized to do so. In \cite{PaperII}, henceforth Paper II, we identified potential patterns in these classes with respect to host star properties. In this work, we return to the orbital architectures of exoplanets to search for trends in their observed eccentricities and inclinations as well as broader measures of dynamical excitation in hopes that they will shed further light on our previously identified classes.

\begin{figure}[!ht]
    \centering
    \includegraphics[width=0.66\textwidth]{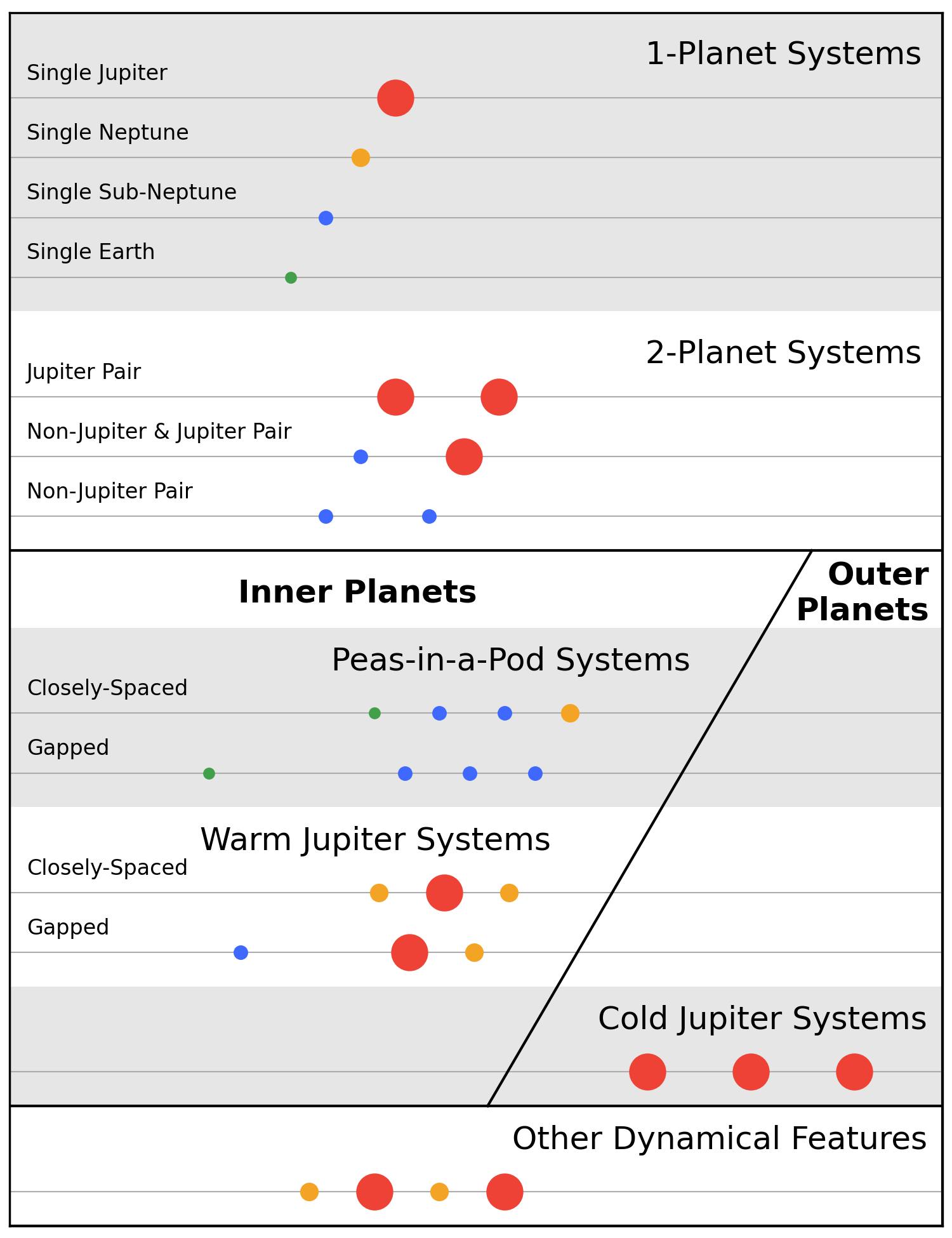}
    \caption{Abridged quick-reference chart for our classification of planetary system architectures presented in Paper I. Each row corresponds to one planetary system, with horizontal spacing corresponding to orbital period on a log scale and point sizes corresponding to planet size. Colors correspond to planet type: jupiters in red, neptunes in yellow, sub-neptunes in blue, and earths in green.}
    \label{fig:quick_reference}
\end{figure}

We provide an abbreviated schematic for our classification framework in Figure \ref{fig:quick_reference}. The core of this classification comes down to three questions for any given system. Does the system have distinct inner and outer planets? Do the inner planets include one or more Jupiters? And do the inner planets contain any gaps with a period ratio greater than 5? This framework allows us to neatly divide the systems with \three inner planets into four categories: the closely-spaced peas-in-a-pod (CPP) systems, the gapped peas-in-a-pod (GPP) systems, the closely-spaced warm jupiter (CWJ) systems, and the gapped warm jupiter (GWJ) systems. Hot jupiters (HJ) can also be considered as their own class, albeit with a small amount of overlap with warm jupiters. We identify additional dynamical features of interest in Paper I, and find that the CPP systems represent a sizable majority (for $N\ge3$). 

The details of our classification system and a catalog of the categories of \three systems are provided in Paper I. In this work, we focus on two features that are particularly important with respect to dynamical excitation: closely-spaced versus gapped systems and systems with outer planets (OP) versus those without. We do not address the subcategories of these systems in this paper because of their small sample sizes.

A few past studies of the connection between eccentricities and system architectures have been made, particularly with respect to multiplicity. For example, \cite{Sagear2025} found higher average eccentricities for single transiting \textit{Kepler} planets orbiting thick disk host stars compared with thin disk stars. More recently, \cite{Gilbert2026} searched for correlations between eccentricities and architectures in \textit{Kepler} compact multiplanet systems. While they detected a correlation between low eccentricity and high multiplicity, they found no other significant correlations within that subpopulation. Our analysis extends past work in this area by considering all confirmed multiplanet systems in the catalog and by dividing them into dynamical classes based on our framework in Paper I.

Of particular interest for our analysis are the gapped systems. These systems often resemble the ubiquitous peas-in-a-pod systems \citep{Weiss2018, Weiss2023} that occur throughout the exoplanet catalog (although such close spacing often occurs in warm-jupiter systems as well), except that there is a gap with a large period ratio that could be large enough to fit an additional planet (in our definition, having a period ratio $>$5). There could, of course, be ``hidden'' planets within these gaps \citep[e.g.,][]{Buchhave2016, Lammers2025}, but any such planet either must be small enough to evade detection or (in the case of transiting systems) must be inclined far enough relative to the other planets to be non-transiting. Either case would be an outlier from the uniformity normally found in peas-in-a-pod systems, so these hidden planets would still be dynamically anomalous. Alternatively, a gap may be real due to an ejected planet or some other dynamical interaction, or simply because the systems did not produce a planet at the gap location. 

A dynamical origin for an apparently gapped system could leave a detectable signature, with the remaining planets being dynamically excited and having systematically larger eccentricities and inclinations \citep{Rasio1996, Juric2008, Lam2024}. Such dynamical excitation would be evidence against the hypothesis that gapped systems have additional small planets below the detection limit, although distinguishing non-transiting planets from ejected planets is a more complicated problem (see Section \ref{sec:discuss}).

In addition to analyzing the distribution of the orbital parameters themselves, we also compute the normalized angular momentum deficit \citep[NAMD; ][]{Raymond13,Turrini20} as a measure of dynamical excitation. This metric is conserved within a planetary system and is a measure of the difference in total angular momentum of a system relative to a hypothetical version of the system with circular, coplanar orbits.
The NAMD provides more robust results across a range of planet sizes and orbital architectures than the orbital parameters alone. This metric would also be useful for identifying potential new classes of exoplanet systems or outlier systems in eccentricity-inclination space, however we do not find any of these in the current dataset (see Section \ref{sec:k65}).

In our analysis, we find that gapped systems do indeed appear to be more dynamically excited than closely-spaced systems, providing further evidence for the existence of the gaps and supporting a different dynamical history for these systems. This differences are supported quantitatively by K-S tests (see Section \ref{sec:results}) for multiple ways to subdivide the dataset, showing it to be a robust result.


We describe the dataset we use in this paper in Section \ref{sec:data}. We describe our procedures for analyzing orbital parameters and computing angular momentum deficits in Section \ref{sec:amd}. We then list the results of our analysis in Section \ref{sec:results} and discuss our findings in Section \ref{sec:discuss}.

\pagebreak

\section{Dataset}
\label{sec:data}

The list of exoplanets that we use in this paper is based on the one we used in Paper II. Our data are drawn from the Planetary Systems Composite Parameters Table \citep{Archive} of the NASA Exoplanet Archive \citep{Christiansen2025}, as of October 2, 2025. Our methods for filtering this list and making corrections to mass and radius values are described in Papers I and II. The final population we consider after this filtering includes 5881 confirmed planets.

In this paper, we additionally require planets to have published eccentricities and/or inclinations (depending on the specific analysis), which are not limits. In addition to further narrowing our catalog, this presents additional complications due to both observational and reporting biases. Furthermore, we consider only multiplanet systems in our analyses. Multiplanet systems are required to compute $\sigma_i$, and we apply the same limitation to our analyses of eccentricities so that we can perform a commensurable analysis on $\sigma_e$.

\subsection{Dataset Biases}
One limitation of our sample selection method is that the observational biases for these orbital parameters are very large. Inclinations are overwhelmingly measured for transiting planets, which are of course limited to inclinations small enough to transit. Meanwhile, eccentricities are measured primary by radial velocity, especially for those measured to high precision. This is partially mitigated because on a population level, we would expect an equipartitioning of angular momentum between the eccentricity and inclination degrees of freedom, especially given the influence of processes such as the Kozai-Lidov mechanism \citep{Kozai1962, Lidov1962}, which can exchange eccentricity and inclination within a system. Because our results in eccentricity space are more robust (as discussed in Section \ref{sec:results}), they can provide clues to the overall state of angular momentum at the population level.

Our dataset is similarly affected by reporting bias.
For most planets, orbital eccentricities in the literature are reported to a precision of 0.01, and disproportionately few planets are listed as having eccentricities of $0<e<0.01$. In many cases, especially for short-period planets, adopted orbital solutions assume $e=0$ \citep[e.g.,][]{Bouchy2010, Anderson2014}. This is sometimes done for statistical reasons, in cases where data does not justify a significantly non-zero eccentricity, to avoid overestimating eccentricities due to the Lucy-Sweeney Bias near $e=0$ \citep{Lucy1971, Zakamska2011}. Other times, this is done for astrophysical reasons: for example, in systems where computed circularization times are low, planets might be expected to be in circular orbits \citep{Adams2006}, and fits may use a zero prior on eccentricity to account for this effect \citep[e.g.,][]{Anderson2012, Vanderburg2017}. 
Together, these factors suggest that the Exoplanet Archive is not ``sensitive'' to eccentricities $e<0.01$. 
This can create further confusion if some planets in a system have a reported $e=0$ and others do not. This could be a result of measurement precision, or different measurement techniques and/or different priors used in the fits. To avoid this issue, we require \textit{all} planets in a system to have a nonzero reported eccentricity.\footnote{Our solar system would barely qualify under this rule with standard reporting practices, as all eight planets have $e>0.005$, which would round up to be reported as nonzero.}

Inclinations do not have this problem with reporting bias, but they have even stronger observational biases. The vast majority of known orbital inclinations for exoplanets are measured for transiting planets, where they can only be detected within a few degrees of 90\textdegree. The exceptions are astrometrically-derived inclinations for very long-period planets \citep[see e.g.][]{Wright2009} and transit timing variation (TTV) measurements with very large uncertainties, as is the case for Kepler-65 e \citep{Mills2019}.

Additionally, inclinations as reported by the Exoplanet Archive are measured relative to the line of sight, not the stellar equator, so the ``true'' inclinations in the sense of the distribution of angular momentum within the system are not available. In this instance, the \textit{dispersion} in inclinations is a more appropriate metric, as that will still encode some information about a system's ``dynamical temperature.'' Nonetheless, a minority of $\sim40$ systems are reported with $\sigma_i=0$, generally due to having reported inclinations of exactly 90\textdegree. Because these may also reflect limitations of measurement precision, we exclude them from our inclination (though not eccentricity) analyses. How we process these data is described in greater depth in Section \ref{sec:amd}.


Finally, requiring all planets in a system to have reported nonzero eccentricities introduces an additional sample-selection bias. Systems are included only if every planet has been analyzed with sufficient precision to measure eccentricity, which preferentially selects systems that have received more extensive follow-up observations or are more amenable to radial-velocity characterization. As a result, our eccentricity sample is not a random subset of the full exoplanet catalog and may be biased toward well-studied or dynamically favorable systems.

Our strategy of examining system architectures using within-system dispersion helps mitigate this effect. Because the dispersion in eccentricity and inclination depends only on the relative values within a system, it is less sensitive to systematic offsets in the absolute measurements. 
As a result, biases that produce a systematic offset in measured parameters will cancel when computing the dispersion, allowing our metric to capture differences in dynamical excitation between systems more robustly than the individual orbital parameters alone.

\subsection{Dataset Description}
In the course of our analysis, we made one manual correction to the data listed by the Exoplanet Archive. Kepler-342 e was listed in the most recent solution as having an inclination of 58.94\textdegree, an impossibility for a transiting planet. However, this appears to be a case of transposed digits, as earlier solutions \citep{Coughlin2016} list an inclination of 85.94\textdegree, so we adopt that value for our analysis \citep{Christiansen2025}.

Table \ref{tab:compare} lists the number of systems in our dataset by data available and by detection method: transit and TTV discoveries only; other discovery methods only (the large majority being radial velocity), and a mix of transit and non-transit discovery methods. The totals for \three planets are also included for comparison. About a third of multiplanet systems have measured nonzero eccentricities, including most radial velocity systems and a small fraction of transiting systems. About two thirds of multiplanet systems have measured inclinations (with nonzero dispersion), which are almost exclusively transiting systems. Only about 10\% of systems have both usable eccentricities and inclinations reported, but that is still enough to compute useful results.

These counts suggest that the selection biases in the sample within each detection method may not be as severe as they appear. Most systems for which we would expect either eccentricity or inclination measurements to be available have them. While the observational biases in the entire catalog of exoplanets are large, the biases in our quantities of interest do not appear to be significantly worse. The measurements available for higher-multiplicity systems are similar in proportion to 2-planet systems, suggesting that the bias with respect to multiplicity is also small.

\begin{table}[htb]
    \centering
    \begin{tabular}{ l | r | r | r | r }
    \hline
    Category          & All Systems & Eccentricities & Inclinations   & Eccentricities   \\
                      & \           & \              & ($\sigma_i>0$) & and Inclinations \\
    \hline
    2+ Planet         &  988 &  298 &  658 & 103 \\
    Transit+TTV Only  &  691 &   93 &  633 &  85 \\
    Non-Transit Only  &  207 &  169 &   13 &  10 \\
    Mixed Detections  &   90 &   36 &   12 &   8 \\
    3+ Planet         &  325 &   80 &  236 &  34 \\
    \hline
    \end{tabular}
    \caption{Number of planetary systems used in our analyses in this paper (top row), broken down by detection method: all transit or transit-timing variations (TTV) discoveries, all non-transit or TTV discoveries, and mixed detection methods. All planets in a system must have eccentricities reported to be nonzero to be included, and all planets in a system must have inclinations reported with nonzero dispersion to be included. The total numbers of \three systems are also tabulated for comparison.}
    \label{tab:compare}
\end{table}

\begin{figure}[]
\centering
\includegraphics[width=0.90\textwidth]{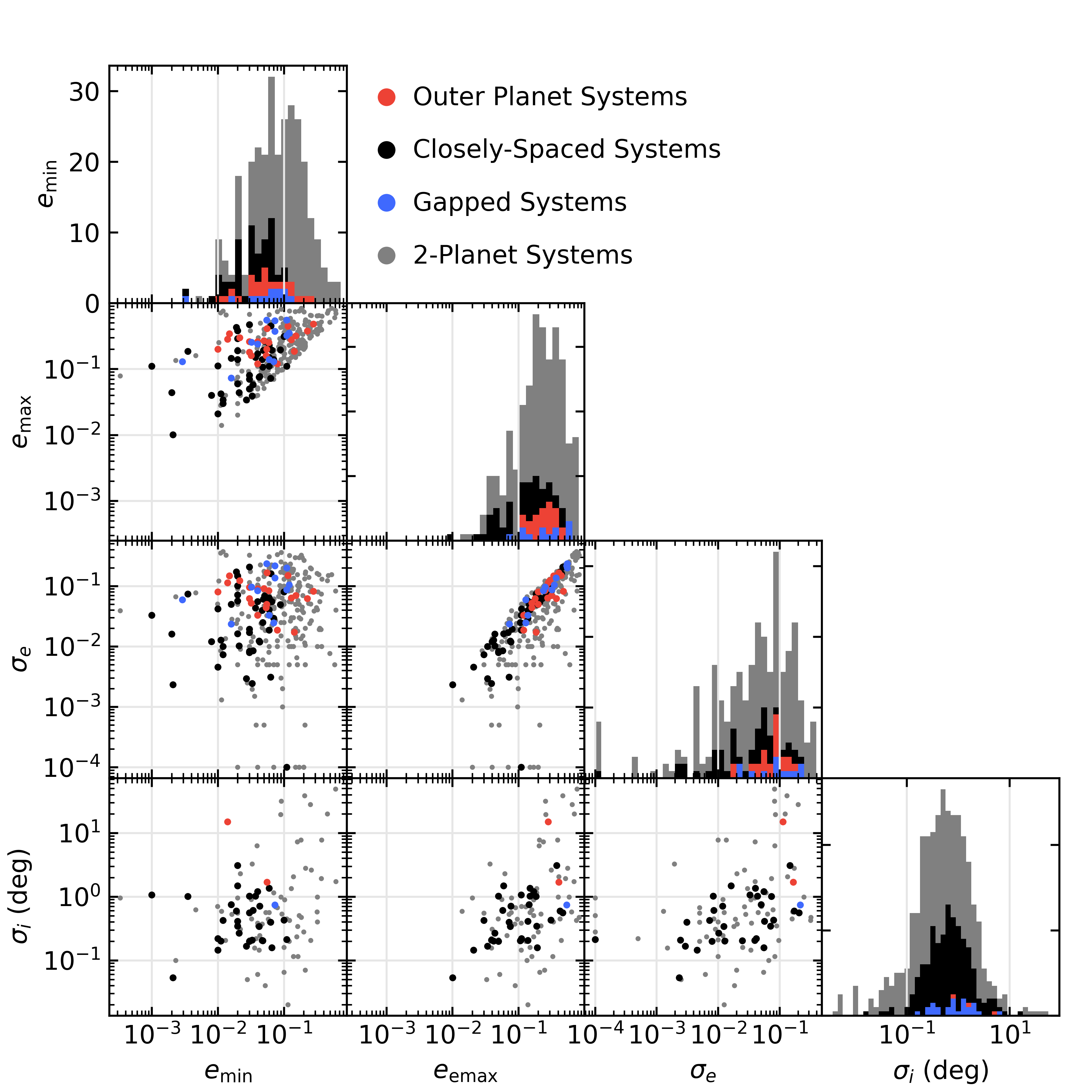}
\caption{Corner plot showing the distributions of minimum eccentricity ($e_{\rm min}$), maximum eccentricity ($e_{\rm max}$), eccentricity dispersion ($\sigma_e$), and inclination dispersion ($\sigma_i$) for our dataset. The top three rows include all systems with usable eccentricity measurements; the bottom right panel includes all systems with usable inclination measurements, and the remainder of the bottom row includes systems with both eccentricity and inclination measurements.} Systems for which $\sigma_e=0$ are plotted as $\sigma_e=10^{-4}$. Results for 2-planet systems are marked in gray. The \three systems are split into systems with detected outer planets (red), gapped systems without outer planets (blue), and closely-spaced systems without outer planets (black).
\label{fig:corner}
\end{figure}

The corner plot in Figure \ref{fig:corner} shows the distribution of four quantities of interest in our dataset: the minimum eccentricity in each system ($e_{\rm min}$), the maximum eccentricity ($e_{\rm max}$), dispersion in eccentricity ($\sigma_e$), and dispersion in inclination ($\sigma_i$). The first three rows plot all systems in our dataset with measured eccentricities. The histogram for $\sigma_i$ includes all systems in our dataset with measured inclinations, while the scatterplots in the bottom row include only systems with both measurements. Note that for some systems, $\sigma_e=0$ even though the individual eccentricities are nonzero because all of the reported eccentricities are identical. We have plotted these systems as $\sigma_e=10^{-4}$ on the logarithmic scale.

For the purpose of our analysis in this paper, we divide our dataset into four categories. The small gray dots are 2-planet systems, which do not have enough data to fully classify. The red dots include all \three systems with detected outer planets, even if their inner planets could be otherwise classified under our framework. Because the dynamical properties of outer planets are so different from inner ones, we consider their presence to be the priority in this case. The black dots are closely-spaced systems without outer planets, and the blue dots are gapped systems without outer planets under our definition. Because of the small sample size of warm-jupiter systems, we do not consider them separately from peas-in-a-pod systems.

Some potentially interesting features in this plot prove to be the result of measurement bias. For example, there appear to be two populations of systems in inclination space: one with a log-normal distribution centered on $\sigma_i\sim0.5$\textdegree\ and a wide range of eccentricities, and one with much higher inclinations and eccentricities. However, these clusters are merely the result of two different measurement techniques. The low-inclination cluster is composed of transit measurements, while the high-inclination cluster is composed mainly of longer-period astrometric measurements.

When we investigated the distributions for each detection method, we found that non-transiting and mixed systems almost always have $\sigma_i>1$\textdegree, while this is uncommon for purely transiting systems, and no purely transiting systems have $\sigma_i>10$\textdegree. In contrast, there is a small tendency towards lower eccentricities in transiting systems, but the detection methods are mostly well-mixed in eccentricity space. We find similar results for our calculations of angular momentum deficits described in Section \ref{sec:amd}.

In eccentricity-space, $e_{\rm min}$ also appears to follow a log-normal distribution centered on $\sim0.05$, but it cuts off at 0.01, while there appears to be a tail at $0<e_{\rm min}<0.01$ with fairly high $e_{\rm max}$. However, this apparent trend is also the result of measurement bias, as discussed above, in that eccentricities are usually only reported to a precision of 0.01, and the tail simply consists of those that are reported to higher precision.

Other features are more robust. The \three systems have systematically lower eccentricities than 2-planet systems (possibly as a result of having many long-period giant planets in the 2-planet sample), and closely-spaced systems have lower eccentricities than the other \three systems, suggesting that these systems are dynamically distinct. In particular, the lower bound in both $e_{\rm max}$ and $\sigma_e$ is an order of magnitude lower for closely-spaced systems than it is for both gapped systems and outer planet systems. In contrast, gapped and outer planet systems appear very similar to each other in distribution, given their different dynamical architectures. (We demonstrate these results quantitatively in Section \ref{sec:results}.)

We also note that $e_{\rm max}$ for outer planet systems ranges from 0.1 to 0.5. Upon closer inspection, out of the 30 measured eccentricities of outer planets in our dataset, only two are smaller than the value of $e=0.057$ for Saturn, which has the largest detectable eccentricity in our own solar system. (Mercury and Mars both have higher eccentricities, but they would be smaller than the detection limit.) This suggests that while our solar system is modestly less regular than average in the architecture of the inner planets (as discussed in Paper I), it appears to be somewhat \textit{more} regular than average in the dynamics of the \textit{outer} planets.

It is also notable that the correlation between $e_{\rm min}$ and $e_{\rm max}$ is surprisingly weak, significantly weaker than the correlations in mass ratio and period ratio that typify the ``peas-in-a-pod'' phenomenon. It is quite common to have planets with low eccentricity and high eccentricity in the same system, suggesting the processes that lead to them may be less uniform than the formation processes that lead to masses and periods.

\section{Angular Momentum Deficit}
\label{sec:amd}

The distribution of orbital eccentricities and inclinations in multiplanet systems is of interest because larger eccentricities and inclinations suggest a more dynamically excited system that could point to a difference in formation processes. 
A common metric to evaluate dynamical excitation for planetary systems is the angular momentum deficit (AMD), defined by \cite{Laskar97,Laskar00} and widely used to characterize orbital excitation in exoplanet systems \citep[e.g.,][]{Laskar2017, Petit2017, He2020, Tamayo2020}. For a single planet, $p$, the AMD is given by
\begin{equation}
    {\rm AMD}_p=m_p\sqrt{GM_*a_p}\left(1-\sqrt{1-e_p^2}\cos i_p\right),
\end{equation}
where $i_p$ is measured from the system's invariable plane. This definition is in contrast to the convention in exoplanet orbit fitting, where inclination is based on the line of sight, and an edge-on system is held to have $i=90$\textdegree. The planet with the greatest angular momentum will contribute the most to the orientation of the system's invariable plane, so we adopt this planet's orbit as a fiducial reference plane. Thus, in this analysis, we adopt a definition of $i_p = i_{\rm los} - i_{\max L}$ for orbital inclination.\footnote{Our calculation of angular momentum for this purpose includes eccentricity in Figure \ref{fig:namd}, where all of the plotted systems have measured eccentricities, but does it not include eccentricity in our other analyses because they include many systems with only measured inclinations.}

AMD$_p$ is the difference between the planet's actual angular momentum and the angular momentum of a circular orbit, coplanar with the invariable plane, and with the same semi-major axis:
\begin{equation}
    L_{\rm circ,p}=m_p\sqrt{GM_*a_p}.
\end{equation}
The total system AMD is the sum over all planets in the system: ${\rm AMD} = \sum_p {\rm AMD}_p$.
Comparing AMD$_p$ for planets in different systems is complicated because the star mass, planet masses, and semi-major axes are all different. Therefore, in this work, we use the normalized angular momentum deficit (NAMD$_p$) \citep{Raymond13,Turrini20}, which converts AMD to a dimensionless number based solely on the shape of the orbit:
\begin{equation}
    {\rm NAMD}_p=\frac{{\rm AMD}_p}{L_{\rm circ,p}} = 1-\sqrt{1-e_p^2}\cos i_p.
\end{equation}

NAMD$_p$ can vary from 0 to 2 for closed orbits.\footnote{It equals 1 for a parabolic orbit, while for a hyperbolic orbit, $1-e^2$ is replaced by $e^2-1$, so that ${\rm NAMD}\rightarrow1-e\cos i$, for large $e$.} For a circular orbit, it is 0 for a 0\textdegree\ inclination and increases with inclination, reaching 1 for a polar orbit and 2 for a retrograde orbit. For an elliptical orbit, NAMD approaches 1 regardless of inclination as eccentricity increases (from below for a prograde orbit and from above for a retrograde orbit).
For an entire planetary system, the total NAMD is computed by adding up the AMD$_p$ of each planet before normalizing (i.e., by weighting each planet's contribution based on its mass):
\begin{equation}
    {\rm NAMD}=\frac{\sum_p{\rm AMD}_p}{\sum_p L_{\rm circ,p}}
\end{equation}

Despite being well-suited to characterize planetary systems, AMD still has limitations based on the inherent limitations of the observations. For small $e$ and $i$, to leading order, NAMD$_p$ can be approximated as:

\begin{equation}
    {\rm NAMD}_p \approx 1-\left(1-\frac{e_p^2}{2}\right)\left(1-\frac{i_p^2}{2}\right),
\end{equation}
for $i_p$ in radians. This expression shows that (in the appropriate limit) eccentricity and inclination contribute equally to NAMD$_p$. In practice, however, measurements of exoplanets are sensitive to different regimes in those orbital elements. Observed exoplanet eccentricities span nearly the full allowed range, from near-circular orbits to highly eccentric orbits reaching up to $e=0.95$ \citep{Dalba2021, Sozzetti2023, Gupta2024} and are typically only reported to a precision of 0.01. Meanwhile, inclination can usually be measured only at $<0.1$ rad, and it is often reported at a precision of $0.01^\circ=0.00017$ rad. Thus, when combining the two quantities, eccentricity will contribute disproportionately to the total NAMD in our dataset.

Additionally, because we can observe the inclinations of most planets only in transit, we are not observing the planet at its maximum distance from the line of sight, because the longitude of ascending node ($\Omega$) is effectively randomized. At the population level, this means that inclinations will be underestimated in magnitude by an average factor of $\langle|\sin \Omega|\rangle=\frac{2}{\pi}$, and their contribution to NAMD$_p$ will be underestimated by an average factor of $\frac{4}{\pi^2}\approx0.4$ for small $i$ and $e$. However, this discrepancy does not have a significant effect on our results.

Because eccentricities and inclinations probe such different regions of the parameter space and are measured for different subsets of objects, in the following section, we consider their impact on NAMD both separately (setting one or the other equal to zero) and together.

\pagebreak

\section{Results}
\label{sec:results}

\subsection{Contribution of Eccentricity (Assuming Coplanar Orbits, $i=0)$}

For planets with measured eccentricities, inclinations are available for only about one third of systems. Because these inclinations are generally small, it is a good approximation at the population level to set all of their inclinations equal to zero and consider the NAMD as defined only by eccentricity, NAMD$_e$, based on the formula,

\begin{equation}
    {\rm NAMD}_{e,p} = 1-\sqrt{1-e_p^2}.
\end{equation}

\begin{figure}[]
\centering
\includegraphics[width=0.90\textwidth]{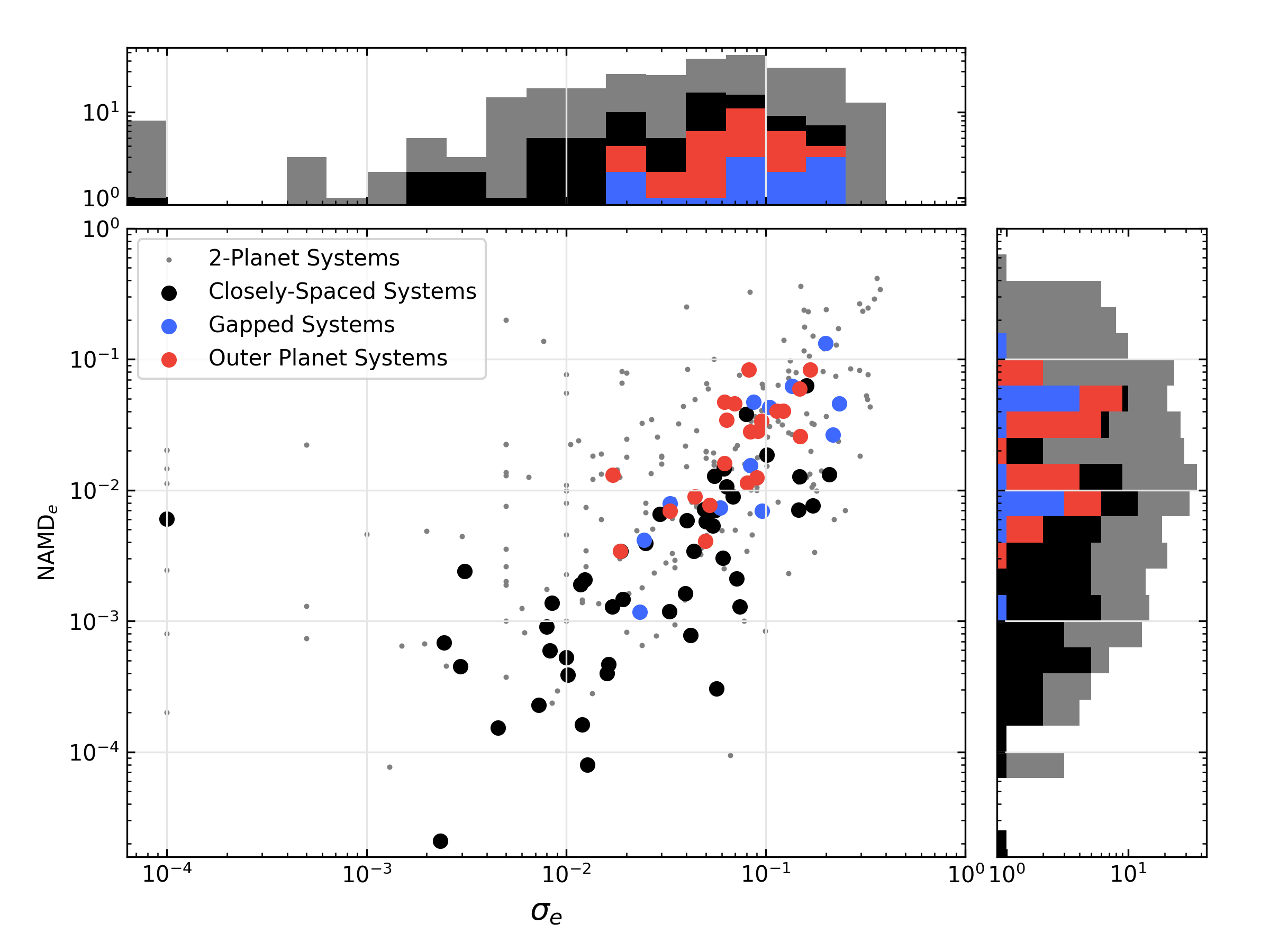}
\caption{Scatterplot of the system angular momentum deficit in eccentricity NAMD$_e$ (calculated by assuming $i=0$) versus dispersion $\sigma_e$ in eccentricity. Histograms of the two quantities are shown on the top and right of the plot.}
\label{fig:namd_ecc}
\end{figure}

We plot the system NAMD$_e$ versus $\sigma_e$ in Figure \ref{fig:namd_ecc}, with systems for which $\sigma_e=0$ again plotted as $\sigma_e=10^{-4}$. As we would expect, the two quantities are correlated. Interestingly, the correlation is significantly better for \three systems than for 2-planet systems. However, this is likely a statistical artifact of system multiplicity. The 2-planet systems are more likely to have non-zero eccentricities with low dispersion than higher-multiplicity systems, which would place them farther to the left in the plot. Otherwise, the distribution is broadly similar between NAMD$_e$ and $\sigma_e$. The 2-planet systems extend to higher values than \three systems; closely-spaced systems extend to lower values then both gapped and outer-planet systems, and gapped and outer-planet systems are distributed similarly to each other.

The outlier \three system at the far left is K2-72, for which all four planets have adopted eccentricities of 0.11.\footnote{In the solution of \citet{Dressing2017}, the eccentricities are similar but not identical (differing at the third significant figure) yet they round to the same value at the two-decimal precision reported in the Exoplanet Archive.} The outlier with the smallest NAMD$_e$ is TRAPPIST-1, whose seven planets have eccentricities measured from 0.00208 to 0.01007 \citep{Gillon2017, Luger2017, Agol2021}. This reveals an additional bias in the dataset: sample-selection bias in which high-interest systems will be measured more precisely, which may be significant given that potentially many systems currently reported with $e=0$ need more precise measurements of their eccentricities to be included in our dataset.

\subsection{Contribution of Inclination (Assuming Circular Orbits, $e=0)$}

The largest portion of planetary systems are measured with only inclinations and not eccentricities. Angular momentum deficits can be estimated for these systems only by assuming circular orbits, and then computing NAMD$_i$ based on the formula
\begin{equation}
    {\rm NAMD}_{i,p} = 1-\cos i_p.
\end{equation}

Because measured inclinations are usually small, this formula risks being particularly inaccurate because it fails to capture the larger expected contribution of eccentricity to NAMD. However, in the case of transiting planets, the contribution of eccentricity may also be small because transiting planet orbits are usually nearly circular, and indeed are often assumed to be circular. Therefore, calculations of NAMD using only inclinations are a valuable comparison.

\begin{figure}[]
\centering
\includegraphics[width=0.90\textwidth]{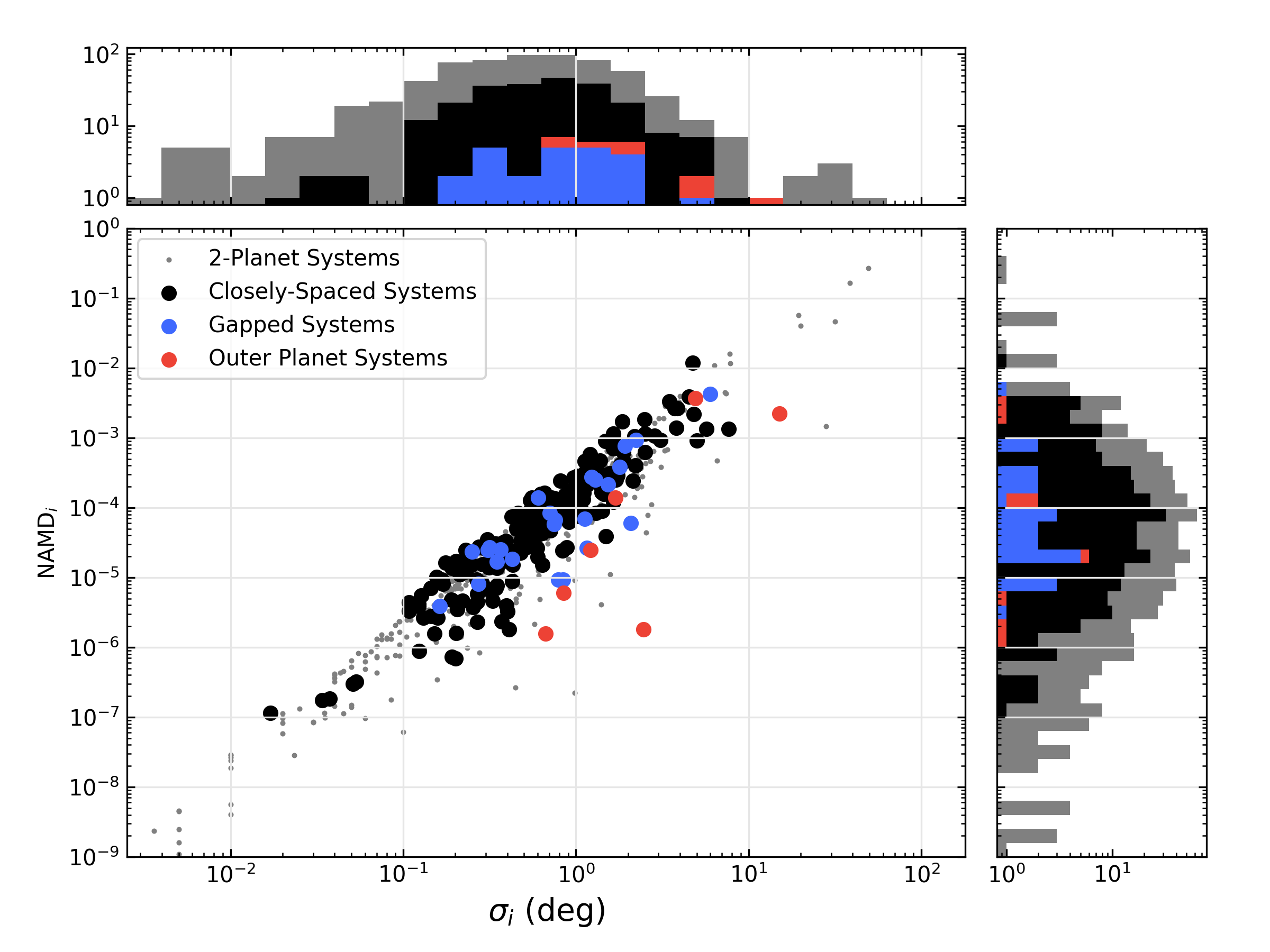}
\caption{Scatterplot of the system angular momentum deficit in inclination NAMD$_i$ (calculated by assuming $e=0$) versus dispersion $\sigma_i$ in inclination. Histograms of the two quantities are shown on the top and right of the plot.}
\label{fig:namd_inc}
\end{figure}

We plot the system NAMD$_i$ versus $\sigma_i$ in Figure \ref{fig:namd_inc}. As expected from the measurement biases, we find ${\rm NAMD}_i \ll {\rm NAMD}_e$, with a median value of 10$^{-4}$ versus 10$^{-2}$. NAMD$_i$ also spans a much wider range, with a few values for radial velocity or astrometrically-measured systems approaching 1 and a few values $<10^{-8}$, reflecting the smaller reported inclinations of many transiting planets.

The outlier toward the lower right among \three systems is 55 Cnc, which is likely an artifact of measurement bias. The adopted inclinations for its planets are all identical (89.73\textdegree) except for Planet e, the smallest and innermost planet, which is inclined at 6\textdegree\ relative to the others. Thus, it contributes significantly to the dispersion in inclination, but contributes unusually little angular momentum to NAMD$_i$. The outlier toward the upper right is Kepler-65 \citep{Chaplin2013, Weiss2024}, the only \three system measured to have a highly-inclined outer jovian companion, with an orbit tilted at least 12\textdegree\ from the plane of the inner planets (best fit 35\textdegree) as measured by an RV+TTV fit.

\begin{figure}[]
\centering
\includegraphics[width=0.90\textwidth]{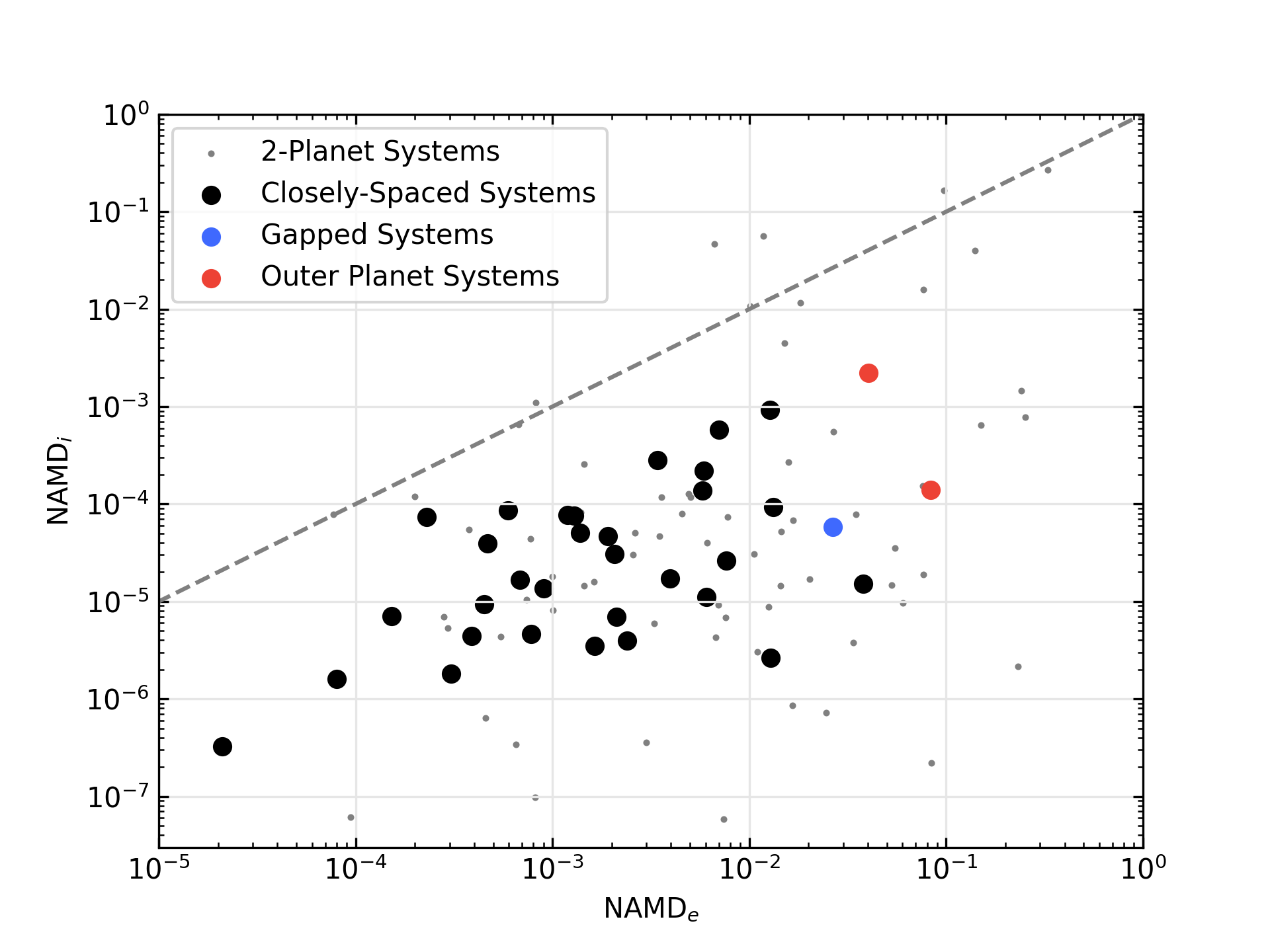}
\caption{Scatterplot comparing the contribution of inclination (NAMD$_i$) versus eccentricity (NAMD$_e$) to total NAMD.}
\label{fig:namd}
\end{figure}

We can quantify the contribution of inclination to NAMD by comparing NAMD$_e$ with NAMD$_i$ directly, which in shown in Figure \ref{fig:namd}. In an unbiased sample, we would expect an equipartitioning such that the contributions of eccentricity and inclination would be of similar magnitude. The generally low measured eccentricities of transiting planets, where available, suggest this might be reflected in our dataset. However, this proves not to be the case. In comparing the two measures, even limited to systems that have both, we find that ${\rm NAMD}_i<0.1\ {\rm NAMD}_e$ for over 80\% of our sample and nearly all of the \three systems.




\subsection{K-S Test Results}

Figures \ref{fig:corner} and \ref{fig:namd_ecc} show qualitatively that gapped systems and systems with detected outer planets both appear to be dynamically excited when compared with closely-spaced inner planets. To quantify this, we perform Kolmogorov-Smirnov (K-S) tests to compare these populations in four of our quantities of interest: NAMD$_e$, $\sigma_e$, NAMD$_i$, and $\sigma_i$. (Too few systems in each category have both eccentricities and inclinations measured to compare the total NAMD.) For this analysis, we consider only \three systems because our classification does not subdivide 2-planet systems based on gap size or period. The results of these tests are listed in Table \ref{tab:ks-test}. The populations being compared include closely-spaced systems (\three systems with no detected outer planet and no inner period-ratio gap larger than 5), gapped systems (\three systems with no detected outer planet and at least one inner period-ratio gap larger than 5), and outer-planet systems (\three systems with at least one detected outer planet), compared pairwise.

\begin{table}[htb]
    \centering
    \begin{tabular}{ l | l  l | l  l }
    \hline
    Category          & \multicolumn{2}{c|}{Overlap counted as Outer} & \multicolumn{2}{c}{Overlap counted as Inner} \\
    \hline
    \  & Significance & Threshold & Significance & Threshold \\
    \hline
    \multicolumn{5}{c}{K-S Tests in NAMD$_e$} \\
    \hline
    Close vs. Gapped  & $p=0.0029$ & 0.0063 & $p=0.0039$ & 0.0083 \\
    Close vs. Outer   & $p=1.2\times10^{-6}$ & 0.0071 & $p=5.9\times10^{-7}$ & 0.0063 \\
    Gapped vs. Outer  & $p=0.75$   & 0.05   & $p=0.45$   & 0.0125 \\
    \hline
    \multicolumn{5}{c}{K-S Tests in $\sigma_e$} \\
    \hline
    Close vs. Gapped  & $p=0.0037$ & 0.0071 & $p=0.0077$ & 0.01 \\
    Close vs. Outer   & $p=0.0014$ & 0.0083 & $p=0.0018$ & 0.01 \\
    Gapped vs. Outer  & $p=0.48$   & 0.0167 & $p=0.40$   & 0.01 \\
    \hline
    \multicolumn{5}{c}{K-S Tests in NAMD$_i$} \\
    \hline
    Close vs. Gapped  & $p=0.80$ & 0.05   & $p=0.53$ & 0.025 \\
    Close vs. Outer   & $p=0.56$ & 0.025  & $p=0.70$ & 0.05  \\
    Gapped vs. Outer  & $p=0.30$ & 0.0071 & $p=0.75$ & 0.05  \\
    \hline
    \multicolumn{5}{c}{K-S Tests in $\sigma_i$} \\
    \hline
    Close vs. Gapped  & $p=0.39$  & 0.0167 & $p=0.21$ & 0.0125 \\
    Close vs. Outer   & $p=0.026$ & 0.0125 & $p=0.12$ & 0.0167 \\
    Gapped vs. Outer  & $p=0.19$  & 0.0063 & $p=0.33$ & 0.0083 \\
    \hline
    \end{tabular}
    \caption{Results in $p$-values of K-S tests comparing three categories of \three exoplanet systems in four of our metrics of interest. Closely-spaced, gapped, and outer planet systems are compared pairwise. For completeness, we consider both our convention in this paper where all systems with outer planets are classified as ``Outer'' (left columns), and our convention in Paper I where systems with outer planets are classified as ``Close'' or ``Gapped'' whenever possible (right columns). For each test, we list a corrected significance threshold beside it based on the Holm-Bonferroni correction ($p<\alpha/(m+1-k)$; \citealt{holm1979}).}
    \label{tab:ks-test}
\end{table}

In this table, we also address the ambiguity in our framework of how to classify outer planets. In the in the left columns, we use the same convention as the remainder of this paper where all systems with outer planets are classified as ``Outer.'' Meanwhile, in the right columns, we adopt our convention from Paper I where systems with outer planets are classified as ``Close'' or ``Gapped'' whenever the categories ``overlap.'' This allows us to probe our Paper I classes directly while still accounting for the effects of the greater angular momentum of giant outer planets. Ultimately, this proves not to have a significant effect on our results.

To address the problem of multiple comparisons in our analysis, we apply the Holm-Bonferroni correction \citep{holm1979} within each pairwise comparison of populations (Close vs. Gapped, Close vs. Outer, and Gapped vs. Outer). This corrections reduces the risk of false positives by lowering the significance threshold for individual tests from $p<\alpha$ to $p_k<\frac{\alpha}{m+1-k}$, where $p_k$ is the $k$-th largest $p$-value out of $m$ tests. (In this case, $m=8$ for each family.) We list the corrected significance threshold for each K-S test alongside its $p$-value.


The K-S tests confirm our qualitative results in terms of eccentricity. The closely-spaced systems are a statistically distinct population from both the gapped systems and the outer-planet systems at high confidence, with significant differences recovered in all of their eccentricity-based comparisons. Meanwhile, the gapped and outer-planet populations are statistically indistinguishable from each other. In contrast, none of our comparisons in inclination are statistically significant, with the populations being either too small or too indistinct from each other to detect any difference. This further supports our prediction that the current sample of inclinations is not adequate to draw any robust conclusions.

\section{Discussion}
\label{sec:discuss}





\subsection{Origins of Gapped Systems}

Planetary systems with large gaps among their inner planets could potentially form by a variety of methods. There may be a planet in the gap that is too small to detect. There may be a planet in the gap that is inclined and non-transiting in an otherwise-transiting system \citep[e.g.,][]{Buchhave2016, Lammers2025}. There may \textit{not} be a planet in the gap because of dynamical instability, such as planet–planet scattering leading to ejection \citep{Rasio1996, Juric2008}. Or there may not be a planet in the gap because some dynamical process during formation prevented a planet from forming there in the first place, as seems to be the case for our solar system's asteroid belt \citep{Izidoro2016}.

We find that gapped systems are dynamically excited compared with closely-spaced systems. This finding provides evidence against the hypothesis that these systems are in fact closely spaced, with one or more planets that is too small to be detected, and instead implies a different dynamical history compared with observed closely-spaced systems. Curiously, systems with detected outer planets appear very similar to gapped systems in these metrics. This similarity may reflect a common dynamical origin; for example, secular perturbations from exterior giant planets could generate gaps and increase gap complexity in inner systems \citep{Livesey2025}.

The specific possible dynamical histories for gapped systems are harder to distinguish. The dynamical excitations that would lead to a planet being ejected would be similar in effect (but different in scale) to the dynamical excitations that would lead to a planet becoming inclined and non-transiting. One possible signature that could differ between these two scenarios is the impact parameters of transits. If the gapped systems, despite being dynamically excited, occupy a narrow range in impact parameter such that $\Delta b=b_{\rm max}-b_{\rm min}$ is systematically $<1$, then we would expect any additional planets to also be transiting. However, in our dataset, we find that $\Delta b$ for both the closely-spaced and gapped systems is uniformly distributed between 0 and 1. 

Another possible indicator is the distribution of the gaps themselves. If gaps are filled with non-transiting planets, then we would expect to see a small proportion of systems with multiple gaps in their inner planets. (We hypothesize in Paper I that gaps between inner and outer planets result from formation processes, possibly involving the dynamics of giant planets.) We do not see any such multiply-gapped inner systems in the current dataset. However, the selection effects mean that the probability of a system being observed as multiply-gapped is low. It would require a minimum of a 5-planet system with non-transiting planets specifically in the second and fourth positions, or a higher-multiplicity system with greater flexibility. For comparison, in our dataset, we see 6 gapped systems with 4 inner planets (which in this hypothesis would be 5-planet systems with 1 non-transiting planet).
We also see 5 gapped systems with 5 inner planets.

The expectation value for the number of multiply-gapped systems is not clear and depends on the probability of whether a planet in a dynamically-excited system is non-transiting. The lack of such systems in the current dataset may be indicative. However, dynamical simulations of systems of these types of systems will likely be needed to distinguish the two hypotheses. Nonetheless, our results still demonstrate a distinct dynamical origin for gapped planetary systems from closely-spaced ones.

\subsection{Kepler-65 and Other Outer Planet Systems}
\label{sec:k65}

In addition to probing dynamical histories, the NAMD could also be useful for identifying new dynamical categories or outlier systems in the data. However, we do not find any new categories or outliers in the current catalog after accounting for observational biases. The best such candidate is Kepler-65. However, after investigating lower-multiplicity systems similar to it, we find that its outlier status is not supported.

When considering only \three systems, Kepler-65 appears to be an outlier, similar to the outlier systems we highlighted in Paper I. Specifically, it is the only known \three system with an outer planet that is highly inclined relative to its inner planets, with a best fit inclination of 35\textdegree\ from the plane of the inner planets. However, when considering measurement biases and comparing it with the data for 2-planet systems, it proves not to be an outlier and may even represent the prototype for a new type of system architecture.

The orbital inclination of a long-period planet can be measured by transits, in the small number of cases that are transiting; by astrometry, which is the source of many of the high-inclination measurements in our dataset; or by transit timing variations (TTVs), which tend to have uncertainties of tens of degrees and will often be consistent with a coplanar orbit even if the best fit is highly inclined. Kepler-65 e has a TTV fit of inclination, in which the $1\sigma$ lower bound on the \textit{mutual} inclination of the planets is 12\textdegree\ \citep{Mills2019}, so it is \textit{not} consistent with coplanar, but it is not the only such case. HAT-P-11 \citep{Bakos2010, Yee2018} is a 2-planet system with a short-period transiting planet and a long-period giant with a period of 9.5 years and an astrometric inclination fit of 144\textdegree$_{-6}^{+5}$, also about 35\textdegree\ away from coplanar.

Both of these examples are also known to be highly eccentric ($e=0.283$ for Kepler-65 e and $e=0.560$ for HAT-P-11 c), as are several other outer planets with measured eccentricities, so equipartitioning arguments suggest that high inclinations would not be surprising. In the limited data available for long-period planets in multi-planet systems, inclinations consistent with coplanar appear to be more common than not, but the non-coplanar examples are enough to predict that a significant fraction of outer planets should have non-coplanar orbits. Future astrometric surveys, especially the upcoming \textit{Gaia} DR4, are expected to yield many more long-period planets with observable inclinations, and thus, they will be able to test this hypothesis much more rigorously.

We also note that outer-planet systems have elevated NAMD$_e$ and eccentricities relative to closely-spaced systems \textit{without} detected outer planets. (We also note that our own outer planets appear unusually regular in comparison.) It is plausible that the processes that create gaps between inner and outer planets, such as the gravitational influence of a massive jupiter-analog, dynamically excite planetary orbits in the same way as the forces that cause gaps within the inner planets, in which case we would expect to see similar results among outer planets. However, as we discuss in Paper I, there is no comparison sample of systems that lack inner-outer gaps (HD 10180 being the only plausible candidate) to compare with these results, so the current dataset is not sufficient to determine if the dynamical excitation is correlated with inner-outer gaps, with outer giant planets in general, or with some other factor.

The other option to examine the influence of giant planets is to look at closely-spaced warm-jupiter systems without detected outer-planets, which limits the available data to systems with giants on closely-spaced orbits. When we examine this subpopulation, we find it is consistent with the closely-spaced peas-in-a-pod systems. However, this is based on a sample size of only 5. Moreover, if we look at outer-planet systems without detected \textit{inner} planets (of which there are 3 in our dataset), they are consistent with the gapped systems, so whether dynamical excitation can be correlated with the presence of giant planets generally remains inconclusive. Again, a larger sample of long-period planets is needed to better understand this population.

\vspace{12pt}

\section*{Acknowledgments}

ARH acknowledges support by NASA under award number 80GSFC24M0006 through the CRESST II cooperative agreement, as well as support from the GSFC Exoplanet Spectroscopy Technologies (ExoSpec).

FCA acknowledges support from the Leinweber Institute for Theoretical Physics (LITP) at the University of Michigan, as well as NSF Grant No. 2508843.

This work was performed in part by members of the Virtual Planetary Laboratory Team, a member of the NASA Nexus for Exoplanet System Science, funded via NASA Astrobiology Program Grant No. 80NSSC18K0829.


We thank the anonymous referee for their assistance in improving the quality of this paper.

This research has made use of the NASA Exoplanet Archive, which is operated by the California Institute of Technology, under contract with the National Aeronautics and Space Administration under the Exoplanet Exploration Program.

This paper makes use of data from the first public release of the WASP data \citep{Butters10} as provided by the WASP consortium and services at the NASA Exoplanet Archive, which is operated by the California Institute of Technology, under contract with the National Aeronautics and Space Administration under the Exoplanet Exploration Program.

This paper makes use of data from the UKIRT microlensing surveys \citep{Shvartzvald17} provided by the UKIRT Microlensing Team and services at the NASA Exoplanet Archive, which is operated by the California Institute of Technology, under contract with the National Aeronautics and Space Administration under the Exoplanet Exploration Program.

This paper makes use of data from the KELT survey, which are made available to the community through the Exoplanet Archive on behalf of the KELT project team.

This paper makes use of data obtained by the MOA collaboration with the 1.8-metre MOA-II telescope at the University of Canterbury Mount John Observatory, Lake Tekapo, New Zealand. The MOA collaboration is supported by JSPS KAKENHI grant and the Royal Society of New Zealand Marsden Fund. These data are made available using services at the NASA Exoplanet Archive, which is operated by the California Institute of Technology, under contract with the National Aeronautics and Space Administration under the Exoplanet Exploration Program.

\software{matplotlib \citep{Hunter:2007},
pandas \citep{mckinney-proc-scipy-2010},
scipy \citep{2020SciPy-NMeth}, 
scikit-learn \citep{scikit-learn}}

\facilities{Exoplanet Archive \citep{Christiansen2025}}

\bibliography{refs}
\bibliographystyle{aasjournal}   

\end{document}